# Functional Integration Approach to Hysteresis

G. Bertotti[a], I.D. Mayergoyz[b], V. Basso[a], and A. Magni[a]

[a]Istituto Elettrotecnico Nazionale Galileo Ferraris, Corso M. d'Azeglio 42, I-10125 Torino, Italy

[b]Electrical & Computer Eng. Dept., University of Maryland, College Park, MD 20742, USA

A general formulation of scalar hysteresis is proposed. This formulation is based on two steps. First, a generating function $g(x)$ is associated with an individual system, and a hysteresis evolution operator is defined by an appropriate envelope construction applied to $g(x)$, inspired by the overdamped dynamics of systems evolving in multistable free energy landscapes. Second, the average hysteresis response of an ensemble of such systems is expressed as a functional integral over the space ***G*** of all admissible generating functions, under the assumption that an appropriate measure $\mu$ has been introduced in ***G***. The consequences of the formulation are analyzed in detail in the case where the measure $\mu$ is generated by a continuous, Markovian stochastic process. The calculation of the hysteresis properties of the ensemble is reduced to the solution of the level-crossing problem for the stochastic process. In particular, it is shown that, when the process is translationally invariant (homogeneous), the ensuing hysteresis properties can be exactly described by the Preisach model of hysteresis, and the associated Preisach distribution is expressed in closed analytic form in terms of the drift and diffusion parameters of the Markovian process. Possible applications of the formulation are suggested, concerning the interpretation of magnetic hysteresis due to domain wall motion in quenched-in disorder, and the interpretation of critical state models of superconducting hysteresis.

## I. INTRODUCTION

The study of hysteresis has been a challenge to physicists and mathematicians for a long time. In physics, hysteresis brings all the conceptual difficulties of out-of-equilibrium thermodynamics [1-5], first of all the fact that we do not know the general principles controlling the balance between stored and dissipated energy in hysteretic transformations [6,7]. In mathematics, on the other hand, the central issue is the formulation of sufficiently general mathematical descriptions grasping the essence of hysteresis beyond the limited interest of ad hoc models [8-11].

In this article, we introduce and discuss a formulation of hysteresis of some generality, inspired by the following situation, often encountered in physical systems. We know that in physics hysteresis is the consequence of the existence of multiple metastable states in the system free energy $F(X)$ (the temperature dependence is tacitly understood), and of the fact that the system may be trapped in individual metastable states for long times. Let us consider the simple case where the state variable $X$ is a scalar quantity and the relevant free energy in the presence of the external field $H$ is $G(X;H) = F(X) - HX$. The metastable states available to the system are represented by $G$ minima with respect to $X$, for which $\partial G/\partial X = 0$, $\partial^2 G/\partial X^2 > 0$. When $H$ is changed over time, the number and the properties of these minima are modified by the variation of the term $-HX$. The consequence is that previously stable states are made unstable by the field action and the system moves to other metastable states through a sequence of Barkhausen jumps. Because the condition $\partial G/\partial X = 0$ is equivalent to $H = \partial F/\partial X$, one can analyze the problem by using the field representation shown in Fig. 1. The response of the system, expressed in terms of $H(X)$, is obtained by traversing the upper and lower envelopes to $\partial F/\partial X$ shown in the figure, the former and the latter applying to increasing and decreasing $H$, respectively. From the physical viewpoint, this construction amounts to assuming that the system, once made unstable by the action of the external field, jumps to the nearest available energy minimum, which means that one excludes the presence of inertial effects, that could aid the system to reach more distant minima.

The method discussed in Section II translates this picture into a well-defined mathematical formulation, based on the following two steps.

(i) Given the time-dependent input $h_t$ and the continuous function $g(x)$, analogous to the free energy gradient $\partial F/\partial X$ of Fig. 1, one associates with them a certain evolution operator $T_{[g]}(h_t)$, which expresses in mathematical terms the envelope construction of Fig. 1 (Section II.1). The function $g$ will be called the *generating function* of $T_{[g]}(h_t)$. The evolution operator acts on a given initial state $s_0$ associated with the initial input $h_0$ and transforms it into the final state $s_t = T_{[g]}(h_t) s_0$.

(ii) Let **G** represent the functional space of all admissible generating functions. Then one constructs a general hysteresis operator as a parallel connection of the collection of operators $T_{[g]}(h_t)$ obtained by varying g over **G**, with appropriate weights described by some measure μ on **G**. Hence, one arrives at the following formulation, in which the overall state $S_t$ describing the collection is expressed in terms of the functional integral

$$S_t = \int_G T_{[g]}(h_t)\, s_0\, d\mu(g) \tag{1}$$

where $s_0$ may depend itself on g (Section II.2).

The generality of the formulation comes from the general nature of the space **G** as well as from general ways of assigning a measure on this space. We will show that several known mathematical descriptions of hysteresis, like the Preisach model [9,12], are particular cases of Eq.(1), and we will discuss some new connections that emerge from the broader perspective offered by the functional integral formulation. A case of particular interest to physics is when Eq. (1) is interpreted as the average hysteresis response of a statistical ensemble of independent systems, each evolving in a different free energy landscape. The space **G** acts then as a probability space and the measure μ describes the probability that an individual system of the ensemble is characterized by a particular generating function g ∈ **G**. In this case, there are situations that can be analytically investigated to a considerable degree of detail, first of all the one where the generating functions $g(x)$ are interpreted as sample functions of a continuous Markovian stochastic process (Section III). In particular, we will show that homogeneous processes give rise to Preisach-type hysteresis, and we will derive explicit analytical expressions for the Preisach distribution as a function of the parameters governing the statistics of the Markovian process.

The results obtained in this article can be important in applications to physics, where randomness due to structural disorder often plays a key role in the appearance of hysteresis effects. The equivalence between Markovian disorder and Preisach-type hysteresis implies that the average system response under small fields is parabolic, a result well known in magnetism under the name of Rayleigh law [12]. In superconducting hysteresis [13], the same equivalence might be of help in the interpretation of critical state models [14,15] in terms of the statistics of the pinning sources acting on Abrikosov vortices, given the equivalence between this class of models and the Preisach model [9,16,17]. Conversely, a limitation of our formulation is the fact that it is based on independent single-degree-of-freedom subsystems, and is thus expected to yield an incomplete description of hysteresis effects arising in systems with more complex internal structures [7,18,19].

## II. MATHEMATICAL DESCRIPTION OF THE MODEL

In this section, we will discuss in more detail the various ingredients defining the structure of the model: input histories, generating functions, admissible states, input-output relationships, stability properties, and finally the functional integration over **G**.

### II.1 Hysteresis in individual systems

Let us consider an individual system characterized by a particular function $g(x)$. The system is acted on by the scalar time-dependent input $h_t$ and generates the scalar output $x_t$, in a way dependent on the function $g(x)$.

*A. Input histories.* We shall consider input histories $h(t)$, $t \geq 0$, such that, at any time, $h_L \leq h(t) \leq h_U$, where $h_L$ and $h_U$ are fixed given fields, delimiting the input range of interest. They will be termed *lower* and *upper saturation field*, respectively. The function $h(t)$ will be assumed to be piece-wise monotone.

*B. Generating functions.* Let us consider a given output interval $[x_L, x_U]$. We say that the function $g(x)$ is an admissible *generating function* associated with the interval $[x_L, x_U]$ if it satisfies the following properties (see Fig. 2):

(i)     $g$ is continuous in $[x_L, x_U]$

(ii)    $g(x_L) = h_L$, $g(x_U) = h_U$                                                                            (2)

(iii)    $h_L < g(x) < h_U$ for any $x$ in the range $x_L < x < x_U$

*C. States.* Any ordered input-output pair $s = (h, x)$ is an admissible state for the system. We will be mainly interested in *equilibrium states*, defined as the states of the form $s = (h = g(x), x)$, with $x_L \leq x \leq x_U$. In other words, an equilibrium state is represented by a point on the generating function. All other states will be generically termed *jump states*. Given the generating function $g(x)$, an equilibrium state is fully described by its output $x$. In this sense, we will often identify an equilibrium state $(h = g(x), x)$ simply by its $x$ value. When the input $h$ is given, the possible equilibrium states under that input are obtained by solving the equation $g(x) = h$. In general, more than one solution will exist. Only the states $s_L = (h_L, x_L)$ and $s_U = (h_U, x_U)$ are unique by definition. They will be termed *lower* and *upper saturation state*, respectively. We will assume that the state of the system before any action is made on it is always an equilibrium state.

*D. Auxiliary functions.* Given the generating function $g(x)$ and the equilibrium state $x_0$, let us introduce the function $h_{[g]}(x; x_0)$, defined as (see Fig. 2):

$$h_{[g]}(x; x_0) = \begin{cases} \min_{[x, x_0]} g & \text{if } x_L \leq x \leq x_0 \\ \max_{[x_0, x]} g & \text{if } x_0 \leq x \leq x_U \end{cases} \quad (3)$$

where the symbols "min" and "max" indicate the minimum and the maximum of $g(x)$ in the specified interval. Function $h_{[g]}$ has the character of a non-decreasing envelope to $g(x)$, more precisely, of an upper envelope for $x \geq x_0$ and a lower envelope for $x \leq x_0$. The inverse of $h_{[g]}(x; x_0)$ will be denoted by $x_{[g]}(h; x_0)$. The mathematical aspects of the connection between $h_{[g]}$ and $x_{[g]}$ are discussed in Appendix A.

*E. System evolution.* Let us introduce the following *evolution operator* $T_{[g]}(h)$, defined in terms of the function $x_{[g]}(h; x_0)$: given the equilibrium state $s_0 = (h_0, x_0)$, with $h_0 = g(x_0)$, and the input value $h$, the state $s$ obtained by applying the input $h$ to $s_0$ is given by the expression

$$s = T_{[g]}(h) \, s_0 = \left( h, x_{[g]}(h; x_0) \right) \quad (4)$$

The evolution of the system is constructed by applying $T_{[g]}(h)$ many times in sequentially, once for each input reversal, as shown in Fig. 3. More precisely, let us suppose that at the initial time $t = 0$ the system is in the equilibrium state $s_0 = (h_0, x_0)$ and let us apply the piece-wise monotone input history $h(t)$. Let us denote by $h_1, h_2, \ldots, h_n$ the sequence of input values at which the input is reversed in the time interval $[0,t]$, and finally let $h_t$ be the current input at the time $t$. Then, the state $s_t$ of the system at the same time is given by

$$s_t = (h_t, x_t) = T_{[g]}(h_t) T_{[g]}(h_n) \ldots T_{[g]}(h_1) s_0 \tag{5}$$

In particular, the output $x_t$ can be expressed in the form

$$x_t = x_{[g]}(h_t; x_n(h_n, \ldots, h_1; x_0)) \tag{6}$$

where $x_n$ - the output value at the last reversal point - depends in general on all past reversal inputs. Note that the evolution is rate-independent, because the state $s_t$ depends only on the current value of the input and on the sequence of past reversal inputs, regardless of the input rate of change.

A relevant aspect of the evolution described by Eq. (5) is that it exhibits *return-point memory* (also called wiping-out property) [19-21]. By this we mean that, given the initial equilibrium state $s_0 = (h_0, x_0)$ and the input extrema sequence $h_0, h_1, h_2, h_1$, with $h_1 > h_0$, $h_0 < h_2 < h_1$, then $T_{[g]}(h_1) T_{[g]}(h_2) T_{[g]}(h_1) s_0 = T_{[g]}(h_1) s_0$ (identical conclusions apply to the case $h_1 < h_0$, $h_0 > h_2 > h_1$). In other words, when the input returns back to the first reversal value $h_1$, the system returns back to the exact same state it occupied when the input first reached the value $h_1$, and the effect of the intermediate input extrema is wiped out. To prove the existence of return-point memory, we begin by remarking that return-point memory is a property of any system whose time evolution satisfies the following properties [19,12]:

(i) the evolution is rate-independent;
(ii) there exists a (partial) ordering relation among the states of the system;
(iii) ordering is preserved during the evolution of the system under the action of ordered input histories.

Property (i) is the direct consequence of the definition of the evolution operator $T_{[g]}(h)$. As property (ii) is concerned, there exists a natural ordering relation deriving from the fact that an equilibrium state is identified by its output value. In fact, given the equilibrium states $s_1 = (h_1, x_1)$ and $s_2 = (h_2, x_2)$, we can simply state that $s_1 \leq s_2$ if $x_1 \leq x_2$ in the usual sense. Finally, property (iii) is the consequence of the theorems of Appendix B, which show that the ordering just defined is preserved by Eq. (5) under the application of ordered input histories. Therefore, return-point memory is indeed a property of Eq. (5).

As discussed in [9, p. 13], return-point memory has the consequence that the final state $s_t$ defined by Eq. (5) is controlled (assuming, for simplicity, that the system is initially in the lower saturation state) by the alternating sequence of dominant extrema $h_{M1}, h_{m1}, h_{M2}, h_{m2}, \ldots$ contained in the full reversal sequence $h_1, h_2, \ldots, h_n$. By this we mean that $h_{M1}$ is the global input maximum in the time interval $[0,t]$, $h_{m1}$ is the global input minimum in the time interval $[t_{M1}, t]$, where $t_{M1}$ is the time at which $h_{M1}$ is reached, and so on.

*F. Stability properties.* They can be conveniently described by introducing the concepts of strong and weak stability. Given the equilibrium states $s_A = (h_A, x_A)$ and $s_B = (h_B, x_B)$, with $x_A < x_B$, we will say that $x_A$ and $x_B$ belong to the same *strongly stable* (*strongly unstable*) interval if $h_A < h_B$ ($h_A > h_B$) and $T_{[g]}(h) s_A = T_{[g]}(h) s_B$ for any input $h$. This definition introduces an equivalence relation among equilibrium states, which generates a partition of the interval $[x_L, x_U]$ into disjoint strongly stable and strongly unstable sub-intervals, possibly reducing to isolated points. Conversely, given the equilibrium states $s_A = (h_A, x_A)$ and $s_B = (h_B, x_B)$, with $x_A < x_B$ and $h_A < h_B$, we will say that the segment $[x_A, x_B]$ is *weakly stable* if $g(x_A) = h_A$, $g(x_B) = h_B$, and $h_A < g(x) < h_B$ for any $x$ in the range $x_A < x < x_B$. Notice that the interval $[x_L, x_U]$ is always weakly stable by definition. When the equilibrium states $s_A = (h_A, x_A)$ and $s_B = (h_B, x_B)$ are such that $x_A < x_B$ and $h_A > h_B$, we will call the segment $[x_A, x_B]$ *weakly unstable* if $g(x_A) = h_A$, $g(x_B) = h_B$, and $h_A > g(x) > h_B$ for any $x$ in the range $x_A < x < x_B$. Each weakly stable or weakly unstable interval will contain in general several strongly stable and strongly unstable sub-intervals. The various possibilities are shown in Fig. 4. One can verify from Fig. 3 that, given any two subsequent reversal points $(h_k, x_k)$ and $(h_{k+1}, x_{k+1})$ associated with a certain input history, the output interval $[x_k, x_{k+1}]$ is always weakly stable. In a sense, the evolution operator $T_{[g]}(h)$ provides a mechanism to select the weakly stable portions of the given generating function. This feature will play an important role in the general formulation of Section II.2 and in the particular cases discussed in Section III.

Stability considerations are important, because the evolution of the system under varying $h$ is reversible inside each strongly stable sub-interval, so that its hysteresis properties are essentially governed by the sequence of jumps occurring from one stable sub-interval to another. A system initially occupying a state inside a weakly unstable interval will never be able to come back to this interval if it ever leaves it. Therefore, only the strongly stable sub-intervals that do not belong to any weakly unstable portion of $[x_L, x_U]$ control the permanent hysteresis properties of the system. Two generating functions possessing the same set of weakly unstable sub-intervals and differing only in their values inside these intervals will give rise to identical hysteresis properties. Considerations of this kind permit one to recognize certain qualitative aspects of hysteresis independent of the details of $g$. For example, Preisach-type hysteresis, briefly discussed in the next sub-section, arises from generating functions containing two strongly stable intervals separated by a weakly unstable part [22].

**II.2 Hysteresis in system ensembles**

Let us now consider an ensemble of systems of the type discussed in the previous sub-section. Each system is identified by a particular generating function $g(x)$ whose domain $[x_L, x_U]$ will be in general different from system to system. We wish to investigate the global hysteresis properties that we obtain when we subject the individual systems of the ensemble to the common input $h_t$ and, roughly speaking, we sum up their responses. The formalism whereby we will carry out this sum in precise mathematical terms is the following. Let us suppose that the response of each individual system is described by some quantity $q_{[g]}$, dependent on the generating function $g$. The ensemble value of that same quantity, say $Q$ (we will use capital letters to denote ensemble properties), will be expressed as a functional integral of the form

$$Q = \int_G q_{[g]} d\mu(g) \qquad (7)$$

Equation (7) is to be interpreted in the following way. The symbol $G$ denotes the functional space of all generating functions satisfying Eq. (2) for some $x_L$ and $x_U$. In general, $x_L$ and $x_U$ will be different for each $g \in G$. One introduces a convenient set (a so-called $\sigma$-algebra) of subsets $A \subset G$ and a positive measure defined over that algebra, $\mu(A) \geq 0$. Then one assumes that there exist elements of the algebra giving rise to values of $q_{[g]}$ inside any arbitrarily small neighborhood of a given value $q_{[g]} = x$, and uses the measure of these subsets to calculate the Lebesgue integral of $q_{[g]}$ over **G**, represented by Eq. (7). To make this loose description mathematically rigorous, one should resort to the language and the methods of measure theory [23,24]. However, it is not the purpose of this paper to go deeper into these mathematical aspects. In the following analysis, it will be sufficient to assume that Eq. (7) does have a precise meaning as a functional integral, and that one knows how to assign the measure $\mu$ in specific cases. In Section III, we will discuss a particular case where one explicitly constructs the measure $\mu$ and expresses the result of the functional integration in a closed analytic form.

As a first step, let us apply Eq. (7) to the definition of ensemble equilibrium states. The main difference with respect to Section II.1 is that we can no longer identify an equilibrium state by its output value. In fact, given the individual output $x$, the corresponding input $g(x)$ may not exist for certain $g$ functions (if $x$ is outside the function domain $[x_L, x_U]$), or may be different from function to function, which is not compatible with the assumption that the entire ensemble is driven by a common input history. In fact, in order to construct a meaningful equilibrium state we must: (i) specify the input value $h_0$; (ii) determine, for each generating function $g \in G$, the set of solutions of the equation $g(x) = h_0$; (iii) for each $g$, select one of these solutions, say $\xi_{[g]}(h_0)$, according to some rule, and build the state $s_{[g]}(h_0) = (h_0, \xi_{[g]}(h_0))$; (iv) construct the ensemble state $S_0$ as

$$S_0 = \int_G s_{[g]}(h_0) \, d\mu(g) = \left( h_0, \int_G \xi_{[g]}(h_0) \, d\mu(g) \right) \tag{8}$$

Equation (8) shows that a great number of possible equilibrium states are associated with a given input $h_0$, as a consequence of the various possible choices for $\xi_{[g]}(h_0)$. Only the lower and upper saturation states are unique, because the equations $g(x) = h_L$ and $g(x) = h_U$ admit just one solution, $x_L$ and $x_U$, for each $g$.

The ensemble evolution is obtained by applying Eq. (7) to Eqs. (4)-(6), that is,

$$S = \int_G T_{[g]}(h) \, s_{[g]}(h_0) \, d\mu(g) = \left( h, \int_G x_{[g]}\big(h; \xi_{[g]}(h_0)\big) \, d\mu(g) \right) \tag{9}$$

$$S_t = (h_t, X_t) = \int_G T_{[g]}(h_t) T_{[g]}(h_n) ... T_{[g]}(h_1) s_{[g]}(h_0) d\mu(g) \tag{10}$$

$$X_t = \int_G x_{[g]}\left(h_t; x_n\left(h_n,...,h_1; \xi_{[g]}(h_0)\right)\right) d\mu(g) \tag{11}$$

Because return-point memory is a property of each individual system, it will also be a property of the ensemble evolution.

The formulation summarized by Eqs. (9)-(11) is rather general and powerful, but it is also quite abstract. It is not obvious how one could possibly determine the measure $\mu$ associated with particular cases and carry out the functional integrals. In this connection, a situation of interest is when one is dealing with a statistical ensemble of independent systems, and one wishes to calculate statistical averages over the ensemble. In that case, Eq. (7) translates into mathematical terms the physical idea that $Q$ represents the sum of all the individual contributions $q_{[g]}$, each weighed by its probability $d\mu(g)$ to occur. Accordingly, $G$ must be endowed with the structure of a probability space: the elements $A$ of the $\sigma$-algebra represent the admissible events that may occur in experiments, the measure $\mu$ satisfies the postulates of probability, and $\mu(A)$ represents the probability of the event $A$.

Probability considerations permit one to express Eqs. (9)-(11) in the following useful form. Let us consider for simplicity the case where the ensemble is initially in the lower saturation state. This eliminates from all equations the complicated dependence on the initial state $\xi_{[g]}(h_0)$ of the individual systems. In particular, Eq. (11) can be written as

$$X_t = \int_G x_{[g]}\left(h_t; x_n(h_n,...,h_1; x_L)\right) d\mu(g) \tag{12}$$

At the end of the previous section, we mentioned the fact that the input reversal sequence $h_1, h_2, ... , h_n$ selects a sequence of weakly stable portions of the generating function. Let us denote by $p(x_t, h_t; h_n,...,h_1) dx_t$ the probability of having a function $g \in G$ such that $g(u) = h_t$ for some $u$ in the interval $[x_t, x + dx_t]$ and such that there exists a sequence of $x$ values, $x_1, x_2, ... , x_n$, for which $g(x_1) = h_1, g(x_2) = h_2 ... , g(x_n) = h_n$ and $[x_L,x_1]$, $[x_1,x_2]$,…, $[x_{n-1},x_n]$, $[x_n,x_t]$ are all weakly stable sub-intervals. Then one can formally write Eq. (12) in the equivalent form

$$X_t = \int_{-\infty}^{\infty} x_t \; p(x_t, h_t; h_n, ..., h_1) \, dx_t \tag{13}$$

The interest of Eq. (13) lies in the fact, discussed in Section III, that the probability density $p$ can be explicitly calculated in the case where the measure $\mu$ is generated by a continuous Markovian stochastic process.

We conclude this section by showing, as an example, when Eqs. (9)-(11) can contain and reproduce other known hysteresis models. We will discuss the Preisach model [9,12]. To this end, let us consider the case where the integral of Eq. (7) is restricted to the subspace $G_P \subset G$ containing the generating functions of the type shown in Fig. 5. The domain $[x_L, x_U]$ is equal to [-1,1] for all functions. Each function is made up of two strongly stable, vertical branches [25], separated by a central, weakly unstable interval. The left branch increases from $h = h_L$ to $h = a$ at $x = x_L = -1$, and the right one increases from $h = b$ to $h = h_U$ at $x = x_U = 1$. One must assume $a > b$ if the central part is to be weakly unstable. Then, let us decompose the space $G_P$ into the equivalence classes $\Lambda_{ab}$ containing all the generating functions characterized by the same $a$ and $b$, and let us express Eq. (12) as an integral over those equivalence classes, that is

$$X_t = \iint_{a>b} \left( \int_{\Lambda_{ab}} x_{[g]}(h_t; x_n(h_n, ..., h_1; x_L)) \, d\mu(g) \right) da \, db \tag{14}$$

As discussed at the end of Section II.1, all generating functions characterized by the same set of weakly unstable intervals and differing only in the values they take inside these intervals give rise to identical hysteresis properties. This means that the function $x_{[g]}$ appearing in Eq. (14) takes the same values for any $g \in \Lambda_{ab}$, so it can be taken out of the integral. We obtain

$$X_t = \iint_{a>b} \gamma_{ab}[h_t]\,\mu(a,b)\,da\,db \tag{15}$$

where $\gamma_{ab}[h_t]$ expresses in simplified operator form the dependence of $x_{[g]}$ on $a$, $b$, and input history, whereas $\mu(a,b)$ represents the measure of the class $\Lambda_{ab}$. It is easy to check through Figs. 3 and 5 that $\gamma_{ab}[h_t] = \pm 1$. In other words, $\gamma_{ab}[h_t]$ is a rectangular-loop operator and the hysteresis model is a weighted superposition of these operators, which means that it is precisely the Preisach model.

In the next section, we will show that the Preisach model can also emerge in a completely different context, when the generating functions $g(x)$ are interpreted as sample functions of a Markovian stochastic process.

## III. MEASURES GENERATED BY STOCHASTIC PROCESSES

The main difficulty of the formulation discussed in the previous section lies in its abstract nature. One needs some tools to generate and manipulate the measure $\mu$ before one can apply the approach to specific situations of interest. In this section, we discuss the case where this issue is addressed by interpreting the generating functions $g(x)$ as sample functions of some stochastic process. We will show that, quite remarkably, the calculation of Eq.(13) can then be reduced to the solution of the *level-crossing problem* (also called exit problem or first-passage-time problem) [26,27] for the stochastic process considered. This will create a direct bridge between two such distant fields as the theory of hysteresis and the theory of stochastic processes, and will permit us to exploit the machinery of level-crossing analysis to derive analytical results on hysteresis. In particular, we will show that homogeneous continuous Markovian processes give rise to Preisach-type hysteresis and we will derive explicit analytical expressions for the associated Preisach distribution $\mu(a,b)$.

### III.1. Markovian processes with continuous sample functions

Let us consider the stochastic process $g_x$. To avoid confusion, we point out that the independent variable $x$ has nothing to do with the real time $t$: it will play the role of a fictitious time to be eventually identified with the system output. We assume that the process is Markovian, that is, its evolution under given initial conditions, say at $x = x_0$, depends on these conditions only and not on the behavior of the process for $x < x_0$. In addition, we assume that the process is a diffusion one, which means that (almost) all its sample functions $g(x)$ are continuous functions of $x$. We will use the letter $h$, with appropriate subscripts, to denote values taken by these sample functions.

In Section II.1.F, we discussed the fact that, given the generating function $g(x)$, any arbitrary input reversal sequence $h_1, h_2, \ldots, h_n$ selects a sequence of weakly stable portions of that function. We will show now that, when $g(x)$ is interpreted as a sample function of $g_x$, weakly stable intervals are naturally and intimately related to the solution of the level-crossing problem for $g_x$. To this purpose, let us consider the interval $[b, a]$ of the $h$ axis and let us select in it the point $h_0$, with $b < h_0 < a$. Let us imagine that we generate a sample function $g(x)$ of the process starting from $(h_0, x_0)$, and that we follow it until it reaches one of the two boundaries, $h = b$ or $h = a$, for the first time (Fig. 6a). The value of $x$ at which $g(x)$ reaches the boundary is a random variable. The problem of determining the statistical properties of this random variable is known in the literature as level-crossing problem or exit problem or first-passage-time problem for the stochastic process [26,27]. Let us restrict the level-crossing analysis to the sample functions that reach the upper boundary $h = a$ first, and let us take the limit $h_0 \to b$, as shown in Fig.6b. If we interpret the function shown in Fig. 6b as a portion of some generating function extending outside the interval $[x_b, x_a]$, we immediately recognize that the interval $[x_b, x_a]$ is weakly stable (see Section II.1.F), because $g(x_b) = b$, $g(x_a) = a$, and $b < g(x) < a$ for any $x$ in the range $x_b < x < x_a$. Therefore, $x_a$ and $x_b$ are admissible reversal outputs that may be encountered under input histories with input reversals at $h = a$ and $h = b$, and the solution of the particular level-crossing problem shown in Fig. 6b is accordingly expected to give direct information about the probability distribution of those reversal outputs.

To analyze in detail the consequences of this idea, let us assume that the particular level-crossing problem of Fig. 6b has been solved, so that we know the conditional probability density $T(a, x_a \mid b, x_b)$ of having a level-crossing event at $x = x_a$ (that is, of having $g(x_a) = a$) conditioned by the fact that $g(x_b) = b$. The function $T$ is defined for $a \geq b$ and is non-anticipating, that is, $T(a, x_a \mid b, x_b) = 0$ for $x_a < x_b$. It obeys the normalization condition

$$\int_{x_b}^{\infty} T(a, x_a \mid b, x_b) \, dx_a = 1 \tag{16}$$

The quantities $x_a$ and $x_b$ are in general random variables. Let us denote by $p_a(x_a)$ and $p_b(x_b)$ their probability distributions. These distributions are not independent, because they must satisfy the equation

$$p_a(x_a) = \int_{-\infty}^{x_a} T(a, x_a \mid b, x_b) \, p_b(x_b) \, dx_b \tag{17}$$

Notice that, because of the Markovian character of the process, Eq. (17) is fully independent of the behavior of the process outside the interval $[x_b, x_a]$. Let us define the space $G$ of Section II.2 as the space containing all those sample functions of the given Markovian process which satisfy the requirements of Eq. (2) for some $x_L$ and $x_U$, that is, $g(x_L) = h_L$, $g(x_U) = h_U$, and $h_L < g(x) < h_U$ for any $x$ in the range $x_L < x < x_U$. In general, $x_L$ and $x_U$ will be random variables, taking different values for each $g \in G$. Then, let us study the evolution of the ensemble described by Eq. (12), assuming that the ensemble is initially in the lower saturation state (i.e., $h(0) = h_L$). Let us denote by $h_1, h_2, \ldots, h_n$ the alternating sequence of dominant input extrema controlling the evolution of the ensemble (this sequence was indicated as $h_{M1}, h_{m1}, h_{M2}, h_{m2}, \ldots$ in Section II.1.E). Let us analyze in some detail what happens when the input $h_t$ increases from $h_L$ up to $h_1$ along the first hysteresis branch, and then decreases from $h_1$ to $h_2$ along the second one. We denote by $x_t$ the output value associated with $h_t$ for a given generating function (see Fig. 7a). The interval $[x_L, x_t]$ is weakly stable for each $g \in G$, so we can apply Eq. (17), with $b = h_L$, $x_b = x_L$, $a = h_t$, $x_a = x_{[g]}(h_t; x_L) = x_t$, $p_b(x_b) = p_L(x_L)$, $p_a(x_a) = p(x_t, h_t)$:

$$p(x_t, h_t) = \int_{-\infty}^{x_t} T(h_t, x_t \mid h_L, x_L) \, p_L(x_L) \, dx_L \,, \quad h_L \leq h_t \leq h_1 \tag{18}$$

The probability distribution $p_L(x_L)$ of the lower saturation output $x_L$ can be chosen at will, it is part of the characterization of the initial state of the ensemble. After that, Eq. (18) permits one to calculate the unknown distribution $p(x_t, h_t)$ on the basis of the known functions $p_L(x_L)$ and $T(h_t, x_t \mid h_L, x_L)$. The distribution $p(x_t, h_t)$ is exactly the function needed in Eq. (13) to calculate the average response of the system, according to the expression

$$X_t(h_t) = \int_{-\infty}^{\infty} x_t \, p(x_t, h_t) \, dx_t \tag{19}$$

Equation (18) can be also used to calculate the probability distribution $p_1(x_1;h_1)$ of the reversal output $x_1$ at $h_1$. We find

$$p_1(x_1;h_1) = \int_{-\infty}^{x_1} T(h_1, x_1 \mid h_L, x_L) \, p_L(x_L) \, dx_L \quad , \quad h_L \leq h_1 \leq h_U \tag{20}$$

Similar considerations apply to the second, decreasing input branch, where $h_t$ decreases from $h_1$ to $h_2$ and $x_t$ accordingly decreases from $x_1$ to $x_2$ (see Fig. 7b). The weakly stable interval to consider is now $[x_t, x_1]$. By applying Eq. (17) to this interval, one obtains

$$p_1(x_1;h_1) = \int_{-\infty}^{x_1} T(h_1, x_1 \mid h_t, x_t) \, p(x_t, h_t; h_1) \, dx_t \quad , \quad h_2 \leq h_t \leq h_1 \tag{21}$$

The main difference with respect to Eq. (18) is that the unknown distribution $p(x_t,h_t;h_1)$ is now inside the integral on the right-hand side of Eq. (21), so Eq. (21) is actually an integral equation for $p(x_t,h_t;h_1)$. It is this difference in the structure of Eqs. (18) and (21) that is responsible for the onset of hysteresis in the average output. The comparison of Fig. 7a with Fig. 7b gives a pictorial illustration of this difference. Although the the probability distributions of $x_L$ and $x_1$ are the same, the level-crossing problems to solve under increasing or decreasing input are different, and therefore give rise to different probability distributions and different average outputs.

The procedure that we have described can be continued to calculate the distribution $p_2(x_2;h_2,h_1)$ of the second reversal output, given by the solution of the integral equation

$$p_1(x_1;h_1) = \int_{-\infty}^{x_1} T(h_1,x_1 \mid h_2,x_2) \, p_2(x_2;h_2,h_1) \, dx_2 \quad , \quad h_L \leq h_2 \leq h_1 \tag{22}$$

and then the distributions $p(x_t,h_t;h_2,h_1)$, $p_3(x_3;h_3,h_2,h_1)$ and so on up to the distribution $p_n(x_n;h_n,\ldots,h_1)$ of the last reversal output. At this point, the probability density $p(x_t,h_t; h_n, \ldots, h_1)$ of the current output at time $t$ is given - depending on whether the current input is increasing or decreasing - by one of the two equations,

$$p(x_t,h_t;h_n,\ldots,h_1) = \int_{-\infty}^{x_t} T(h_t,x_t \mid h_n,x_n) \, p_n(x_n;h_n,\ldots,h_1) \, dx_n \quad , \quad h_n \leq h_t \leq h_{n-1}$$

(23)

$$p_n(x_n;h_n,\ldots,h_1) = \int_{-\infty}^{x_n} T(h_n,x_n \mid h_t,x_t) \, p(x_t,h_t;h_n,\ldots,h_1) \, dx_t \quad , \quad h_{n-1} \leq h_t \leq h_n$$

and the corresponding average output is

$$X_t(h_t;h_n,\ldots,h_1) = \int_{-\infty}^{\infty} x_t \, p(x_t,h_t;h_n,\ldots,h_1) \, dx_t \tag{24}$$

By the analysis just concluded, we have reduced the original functional integral over $G$ (Eq. (12)) to a chain of integrals and integral equations (Eqs. (20), (22), (23)), dependent on the saturation distribution $p_L(x_L)$ (arbitrarily chosen) and the transition density $T(a,x_a \mid b,x_b)$. The central problem is then the calculation of $T(a,x_a \mid b,x_b)$ for a given process.

**III.2. Homogeneous processes**

Particularly simple and interesting results are obtained when the statistical properties of the process considered are translationally invariant with respect to $x$, that is, when the process is homogeneous with respect to $x$. In fact, in this case it is not necessary to determine the complete function $T(a,x_a \mid b,x_b)$ in order to predict the hysteresis properties of the ensemble. To clarify this point, let us come back to the first of Eqs. (23). Because of the assumed homogeneity of the process, $T(a,x_a|b,x_b) = T(a,x_a - x_b|b,0)$. Therefore

$$p(x,h;h_n,...,h_1) = \int_{-\infty}^{x} T(h, x - x_n | h_n, 0) \, p_n(x_n; h_n,...,h_1) \, dx_n \qquad (25)$$

where we have dropped for simplicity the $t$ subscript in $x$ and $h$. According to Eq. (24), the average system response is obtained by multiplying both members of Eq. (25) by $x$ and by integrating over $x$. By expressing $x$ as $x = (x-x_n)+x_n$ and by rearranging the appropriate integrals on the right-hand side, we obtain

$$X(h;h_n,...,h_1) = X_n + X_T(h|h_n) \, , \quad h \geq h_n \qquad (26)$$

where

$$X_T(a|b) = \int_0^{\infty} u \, T(a,u|b,0) \, du \, , \quad a \geq b \qquad (27)$$

and

$$X_n = \int_{-\infty}^{\infty} x_n \, p_n(x_n; h_n,...,h_1) \, dx_n \qquad (28)$$

When the second of Eqs. (23) is the relevant equation, by perfectly similar considerations one obtains

$$X(h;h_n,...,h_1) = X_n - X_T(h_n|h) \, , \quad h \leq h_n \qquad (29)$$

We see that the hysteresis properties of the system are fully controlled by the first moment of $T$ only, given by Eq. (27).

Equation (26) shows that the shape of a generic ascending hysteresis branch starting from the reversal field $h_n$ is the same regardless of the past input history. The influence of past history is summarized in the value of $X_n$ (Eq. (28)), and the branches generated by different histories differ by a mere shift along the $X$ axis. The same is true for descending branches (Eq. (29)). The importance of this result lies in the fact that it implies the validity of the so-called congruency property [9]. It is known that return-point memory (built in the description from the beginning, see Section II.1 E) and congruency represent the necessary and sufficient conditions for the description of a given hysteretic system by the Preisach model [27,9,12]. Therefore, we conclude that the hysteresis generated by a homogeneous, diffusion Markovian process is of Preisach type. The process is fully described by the function $X_T(a|b)$ (Eq. (27)), which is nothing but the Everett function associated with the Preisach description. The function $X_T(a|b)$ represents the average value of $x$ at which the generating function crosses the level $h = a$ for the first time, starting at $x = 0$ from the initial level $h = b$ (see Fig. 6b). The description of hysteresis is reduced to the solution of this particular level-crossing problem for the stochastic process.

Remarkably, this solution can be worked out in closed analytical form. To this end, let us start from the description of the process in terms of its Ito stochastic differential equation [26]

$$dh = A(h)\, dx + B(h)\, dW_x \qquad (30)$$

where $dW_x$ represents the infinitesimal increment of the Wiener process $W(x)$, $A$ and $B$ are independent of $x$ because of the assumed homogeneity of the process, and $x$ plays the role of time. The statistics of the process are fully described by the transition density $P(h,x|h_0,x_0)$, giving the probability density that a sample function of the process take the value $h$ at the position $x$, conditioned to the fact that it takes the value $h_0$ at $x = x_0$. The Fokker-Planck equation for the transition density associated with Eq. (30), $P(h,x|h_0) = P(h,x|h_0,0)$, is

$$\frac{\partial}{\partial x}P(h,x|h_0) + \frac{\partial}{\partial h}\left[A(h)P(h,x|h_0)\right] - \frac{1}{2}\frac{\partial^2}{\partial h^2}\left[B^2(h)P(h,x|h_0)\right] = 0 \qquad (31)$$

As discussed before, the situation of interest is the one depicted in Fig. 6. The process starts, at $x = x_0 = 0$, from $h = h_0$. We wish to determine the statistics of the $x$ value at which the process reaches the level $h = a$ for the first time, in the limit $h_0 \to b$. This is obtained by solving Eq. (31) under the initial condition $P(h,0/h_0) = \delta(h - h_0)$, together with the assumption of absorbing boundary conditions at $h = a$ and $h = b$, and then by taking the limit $h_0 \to b$. The mathematical details of the analysis are discussed in Appendix C. The solution for $X_T(a|b)$, expressed in terms of the function

$$\psi(u) = \exp\left[-2\int_0^u \frac{A(u')}{B^2(u')}\,du'\right] \qquad (32)$$

reads

$$X_T(a\mid b) = \frac{2}{K_{[b,a]}}\int_b^a \left[\int_b^u \psi(u')du'\right]\left[\int_u^a \psi(u')du'\right]\frac{du}{B^2(u)\psi(u)} \qquad (33)$$

where

$$K_{[b,a]} = \int_b^a \psi(u)\,du \qquad (34)$$

**III.3. Preisach distribution associated with a given homogeneous process**

The quantity $X_T(a|b)$ given by Eq. (33) coincides with the Everett function of the Preisach model associated with the homogeneous stochastic process. Therefore, as discussed in [9], the Preisach distribution $\mu(a,b)$ is given by

$$\mu(a,b) = -\frac{1}{2}\frac{\partial^2 X_T(a\mid b)}{\partial a\,\partial b} \qquad (35)$$

By deriving Eq. (33), one finds

$$\mu(a,b) = 2\frac{\psi(a)\psi(b)}{K_{[b,a]}^3}\int_b^a\left[\int_b^u \psi(u')du'\right]\left[\int_u^a \psi(u')du'\right]\frac{du}{B^2(u)\psi(u)} \tag{36}$$

that is, taking into account Eq. (33),

$$\mu(a,b) = \frac{\psi(a)\psi(b)}{K_{[b,a]}^2} X_T(a|b) \tag{37}$$

Let us calculate the Preisach distribution associated with some typical stochastic processes.

*A. Wiener process.* The Wiener process is described by $A(h) = 0$, $B(h) = 1$ (see Eq. (30)). Therefore, we obtain from Eqs. (32) and (34),

$$\begin{aligned}\psi(u) &= 1 \\ K_{[b,a]} &= a - b\end{aligned} \tag{38}$$

By inserting these expressions into Eqs. (33), (36), or (37) we find

$$\begin{aligned}\mu(a,b) &= \frac{1}{3} \\ X_T(a|b) &= \frac{1}{3}(a-b)^2\end{aligned} \tag{39}$$

The Preisach distribution is simply a constant and all hysteresis branches are parabolic (Fig. 8).

*B. Wiener process with drift.* By this, we mean the case where by $A(h) = 1/2\xi$, $B(h) = 1$, with $\xi > 0$. We have

$$\psi(u) = \exp(-u/\xi)$$
$$K_{[b,a]} = \exp(-b/\xi) - \exp(-a/\xi) \tag{40}$$

The Preisach distribution and $X_T(a|b)$ are given by

$$\mu(a,b) = \frac{x \operatorname{cth} x - 1}{\operatorname{sh}^2 x}$$
$$X_T(a|b) = 4(x \operatorname{cth} x - 1) \quad , \quad x = \frac{a-b}{2\xi} \tag{41}$$

The Preisach distribution depends on the difference ($a$-$b$) only, and tends to the value 1/3 when ($a$-$b$) $\to$ 0, in agreement with Eq. (38). Typical hysteresis branches calculated from Eq. (41) are shown in Fig. 9.

*C. Ornstein-Uhlenbeck process.* In this case, $A(h) = -h/\xi$, $B(h) = 1$, with $\xi > 0$. We find

$$\psi(u) = \exp(u^2/\xi)$$
$$K_{[b,a]} = a\,\Phi\!\left(\frac{1}{2},\frac{3}{2};\frac{a^2}{\xi}\right) - b\,\Phi\!\left(\frac{1}{2},\frac{3}{2};\frac{b^2}{\xi}\right) \tag{42}$$

where $\Phi(a,c;x)$ is the confluent hypergeometric function. The Preisach distribution and $X_T(a|b)$ are obtained by inserting these expressions into Eqs. (33), (36), or (37).

## IV. CONCLUSIONS

The formulation developed in the previous sections is general enough to offer various possibilities for further studies and applications. From the mathematical viewpoint, the basic issue is the role of return-point memory in the functional integration approach developed in Section II. We know that return-point memory is inherent in the formulation, and it is natural to ask under what additional conditions, if any, an arbitrary scalar hysteretic system exhibiting return-point memory can be described through Eqs. (10)-(13), by choosing appropriately the space *G* and the measure *μ*. For the moment, we do not have a general answer to this basic question.

From the physical viewpoint, the results obtained in the case where the measure is associated with a stochastic process are of direct interest to all those situations where some dominant degree of freedom, say *X*, evolves in a random free energy landscape, and the associated dynamics are overdamped. By this we mean that *X* obeys an equation of the form

$$\gamma \frac{dX}{dt} = H(t) - \frac{\partial F(X)}{\partial X} \tag{43}$$

where *H*(*t*) is the time-dependent driving field, *F*(*X*) is the free energy of the system and $\gamma > 0$ is some typical friction constant. Under small enough field rates, the solutions of Eq. (43) - once expressed in terms of *H* as a function of *X* - precisely approach the behavior shown in Fig. 1 [12,29], so that our formulation can be directly applied, if one knows the statistical properties of the free energy gradient $\partial F/\partial X$. A particularly important example is the motion of magnetic domain walls in ferromagnets, where Eq. (43) often provides a good physical description, and various forms of structural disorder (point defects, dislocations, grain boundaries, etc.) are responsible for the random character of $\partial F/\partial X$. There are a series of classical papers in the literature [30,31], where the domain wall picture has been applied to the prediction of coercivity and magnetization curve shapes, starting from some assumption about the properties of *F*(*X*). Equations (33) and (36) provide a general solution for the case where the process $\partial F/\partial X$ is Markovian, continuous, and homogeneous. In particular, the proven equivalence of Markovian disorder to the Preisach model gives a sound statistical interpretation of the latter in terms of stochastic dynamics in quenched-in disorder. In this respect, the extension of the analysis of Section III to non-Markovian and/or non-homogeneous processes would be of definite interest, as a way to provide quantitative predictions of hysteresis features under more realistic conditions and to indicate in what direction one should generalize the Preisach model in order to improve the macroscopic description of hysteresis generated by various forms of structural disorder.

# APPENDIX A. PROPERTIES OF $h_{[g]}(x;x_0)$ AND $x_{[g]}(h;x_0)$

The function $h_{[g]}(x;x_0)$ defined by Eq. (3) is continuous and non-decreasing with respect to $x$ and it is continuous with respect to $x_0$. Its main properties derive from the following theorem.

*Theorem* A.1. Given any $h \in [h_L, h_U]$, there exists at least one equilibrium state $x \in [x_L, x_U]$, such that $h_{[g]}(x;x_0) = h$.

*Proof.* The theorem holds by definition for $h = g(x_0)$, $h = h_L$, and $h = h_U$. Then, let us consider the case $g(x_0) < h < h_U$. Let us introduce the set

$$U_{[g]}(h, x_0) = \{x \in [x_0, x_U] : g(x) = h\} \tag{A.1}$$

Because of the continuity of $g$, $U_{[g]}$ is closed and will therefore contain its minimum $x_m$. This is the smallest $x \in [x_0, x_U]$ for which $g(x) = h$. Because $g(x_0) < h$, then $g(x) < h$ for any $x \in [x_0, x_m)$, i.e., $\max\{g(u) : u \in [x_0, x_m]\} = g(x_m) = h$. Therefore, the equilibrium state $x_m$ is such that $h_{[g]}(x_m; x_0) = h$. The proof of the case $h_L < h < g(x_0)$ is analogous. The only difference is that one must consider the maximum $x_M$ of the set

$$L_{[g]}(h, x_0) = \{x \in [x_L, x_0] : g(x) = h\} \tag{A.2}$$

Following the results of Theorem A.1, let us introduce the function $x_{[g]}(h;x_0)$ defined as follows:

$$x_{[g]}(h; x_0) = \begin{cases} \max L_{[g]}(h, x_0), & \text{if } h_L \leq h \leq g(x_0) \\ \min U_{[g]}(h, x_0), & \text{if } g(x_0) \leq h \leq h_U \end{cases} \tag{A.3}$$

The function $x_{[g]}(h;x_0)$ is the inverse of $h_{[g]}(x;x_0)$. In fact, according to Theorem A.1, $h_{[g]}(x_{[g]}(h;x_0);x_0) = h$ (see Fig. 2). It monotonically increases with $h$ and, as a rule, it is not continuous in $h$. However, it is continuous in $x_0$, because both $\max L_{[g]}$ and $\min U_{[g]}$ are continuous in $x_0$, as a consequence of the continuity of $g$. Notice that the graph of $x_{[g]}$ consists uniquely of equilibrium states, that is, of points of the generating function $g(x)$.

# APPENDIX B. ORDERING PROPERTIES OF $T_{[g]}(h)$

As discussed in Section II.1, given the equilibrium states $s_1 = (h_1, x_1)$ and $s_2 = (h_2, x_2)$, we say that $s_1 \leq s_2$ if $x_1 \leq x_2$ in the usual sense. The set of equilibrium states is totally ordered with respect to this relation. Before considering the theorems deriving from the existence of this ordering relation, let us prove the following four lemmas, involving the sets $U_{[g]}$ and $L_{[g]}$ defined by Eqs.(A.1) and (A.2).

*Lemma* B.1. Given $x_A \leq x_B$ and $h_L \leq h \leq \min\{g(x_A), g(x_B)\}$, then $\max L_{[g]}(h, x_A) \leq \max L_{[g]}(h, x_B)$.

In fact, $L_{[g]}(h, x_B) = L_{[g]}(h, x_A) \cup L_{AB}$, where $L_{AB} = \{x \in (x_A, x_B] : g(x) = h\}$. $L_{AB}$ is empty or contains elements that are all greater than any element of $L_{[g]}(h, x_A)$. In both cases, the lemma is proven.

*Lemma* B.2. Given $x_A \leq x_B$ and $h_U \geq h \geq \max\{g(x_A), g(x_B)\}$, then $\min U_{[g]}(h, x_A) \leq \min U_{[g]}(h, x_B)$.

In fact, $U_{[g]}(h, x_A) = U_{[g]}(h, x_B) \cup U_{AB}$, where $U_{AB} = \{x \in [x_A, x_B) : g(x) = h\}$. $U_{AB}$ is empty or contains elements that are all smaller than any element of $U_{[g]}(h, x_B)$. In both cases, the lemma is proven.

*Lemma* B.3. Given $x_A \leq x_B$ and $g(x_A) \geq h \geq g(x_B)$, then $\max L_{[g]}(h, x_A) \leq \min U_{[g]}(h, x_B)$.

In fact, given any $x \in L_{[g]}(h, x_A)$ and $x' \in U_{[g]}(h, x_B)$, $x \leq x_A \leq x_B \leq x'$. This will hold in particular for $x = \max L_{[g]}(h, x_A)$ and $x' = \min U_{[g]}(h, x_B)$, which proves the lemma.

*Lemma* B.4. Given $x_A \leq x_B$ and $g(x_A) \leq h \leq g(x_B)$, then $\min U_{[g]}(h, x_A) \leq \max L_{[g]}(h, x_B)$.

In fact, the set $Z_{AB} = \{x \in [x_A, x_B] : g(x) = h\}$ is not empty, which means that $\min U_{[g]}(h, x_A) = \min Z_{AB}$ and $\max L_{[g]}(h, x_B) = \max Z_{AB}$. Because $\min Z_{AB} \leq \max Z_{AB}$, the lemma is proven.

*Theorem* B.1. Given the equilibrium states $s_A$ and $s_B$, $s_A \leq s_B$, then $T_{[g]}(h)\, s_A \leq T_{[g]}(h)\, s_B$ for any $h \in [h_L, h_U]$.

*Proof.* Let us express the states as $s_A = (h_A = g(x_A), x_A)$ and $s_B = (h_B = g(x_B), x_B)$, with $x_A \leq x_B$. According to Eq. (4), the application of $T_{[g]}(h)$ changes $x_A$ and $x_B$ into $x_{[g]}(h;x_A)$ and $x_{[g]}(h;x_B)$. By applying Lemmas B.1-B.4 to the definition of $x_{[g]}(h;x_0)$ (Eq. (A.3)), one finds that $x_{[g]}(h;x_A) \leq x_{[g]}(h;x_B)$ for any $h \in [h_L, h_U]$.

*Theorem* B.2. Given the equilibrium states $s_A$ and $s_B$, $s_A \leq s_B$, and the inputs $h_A$ and $h_B$, $h_A \leq h_B$, then $T_{[g]}(h_A) s_A \leq T_{[g]}(h_B) s_B$.

*Proof.* Let us express the states as $s_A = (h_A = g(x_A), x_A)$ and $s_B = (h_B = g(x_B), x_B)$, with $x_A \leq x_B$. From Theorem B.1, we have that $x_{[g]}(h_A;x_A) \leq x_{[g]}(h_A;x_B)$. On the other hand, $x_{[g]}(h_A;x_B) \leq x_{[g]}(h_B;x_B)$, because of the monotonicity of $x_{[g]}(h;x_0)$ with respect to $h$. Therefore, $x_{[g]}(h_A;x_A) \leq x_{[g]}(h_B;x_B)$.

*Theorem* B.3. Let us consider the initial equilibrium states $s_A(0)$ and $s_B(0)$, $s_A(0) \leq s_B(0)$, and let us apply to them the two ordered input histories $h_A(t) \leq h_B(t)$. Then, at any subsequent time $t > 0$, $s_A(t) \leq s_B(t)$.

*Proof.* Let us consider the evolution of the two states $s_A(t)$ and $s_B(t)$ first from $t = 0$ up to the time of the first reversal of $h_A(t)$ or $h_B(t)$, then from this time to the time of the second reversal of $h_A(t)$ or $h_B(t)$, and so on. By considering that initially $s_A(0) \leq s_B(0)$ and by applying Theorem B.2, we find that order is preserved in the first interval and that the states at the end of the interval are still ordered. This permits one to conclude that order is preserved also in the second interval, and so on.

**APPENDIX C. SOLUTION OF LEVEL-CROSSING PROBLEM**

The probability current associated with Eq. (31) is:

$$J(h,x|h_0) = A(h)P(h,x|h_0) - \frac{1}{2}\frac{\partial}{\partial h}\left[B^2(h)P(h,x|h_0)\right] \quad \text{(C.1)}$$

The rate at which the process leaves the interval [b,a] starting from h = $h_0$ at x = 0 is obtained by integrating Eq. (31) over h. One obtains

$$-\frac{\partial}{\partial x}\left[\int_b^a P(h,x|h_0)\,dh\right] = J(a,x|h_0) - J(b,x|h_0) \tag{C.2}$$

Equation (C.2) shows that the probability current at the boundaries is just proportional to the probability density that a level-crossing event takes place at the position x. As mentioned before, we are interested in level-crossing through the upper boundary h = a, described by the probability current $J(a,x|h_0)$. According to Eq. (C.1), the functional dependence of $J(a,x|h_0)$ on x and $h_0$ is the same as that of $P(a,x|h_0) = P(a,x|h_0,0) = P(a,0|h_0,-x)$. This means that $J(a,x|h_0)$ obeys the backward equation [24]

$$\frac{\partial}{\partial x}J(a,x|h_0) - A(h_0)\frac{\partial}{\partial h_0}J(a,x|h_0) - \frac{1}{2}B^2(h_0)\frac{\partial^2}{\partial h_0^2}J(a,x|h_0) = 0 \tag{C.3}$$

The total probability, $\pi_{[b,a]}(h_0)$, that the process leave the interval [b,a] through h = a is given by the expression

$$\pi_{[b,a]}(h_0) = \int_0^\infty J(a,u|h_0)\,du \tag{C.4}$$

By integrating Eq.(C.3) from x = 0 to x = ∞, and by taking into account that $J(a,0|h_0) = J(a,\infty|h_0) = 0$, one finds that $\pi_{[b,a]}(h_0)$ satisfies the differential equation

$$\frac{1}{2}B^2(h_0)\frac{d^2\pi_{[b,a]}}{dh_0^2} + A(h_0)\frac{d\pi_{[b,a]}}{dh_0} = 0 \tag{C.5}$$

with the boundary conditions

$$\pi_{[b,a]}(a) = 1$$
$$\pi_{[b,a]}(b) = 0 \tag{C.6}$$

deriving from the fact that the process will certainly cross the boundary $h = a$ if it starts from that same level, that is, from $h_0 \to a$, whereas it will never reach $h = a$ if it starts from $h_0 \to b$, because in that case it will certainly cross the boundary $h = b$ first. The solution of Eq. (C.5) is then

$$\pi_{[b,a]}(h_0) = \frac{1}{K_{[b,a]}} \int_b^{h_0} \psi(u) \, du \tag{C.7}$$

where $\psi(u)$ and $K_{[b,a]}$ are given by Eq. (32) and Eq. (34), respectively.

The probability density $p_{[b,a]}(x|h_0)$ that the process reach the boundary $h = a$ at the position $x$ is given by

$$p_{[b,a]}(x|h_0) = \frac{J(a,x|h_0)}{\pi_{[b,a]}(h_0)} \tag{C.8}$$

and the mean value $x_{[b,a]}(h_0)$ of the level-crossing position is

$$x_{[b,a]}(h_0) = \int_0^\infty u \, p_{[b,a]}(u|h_0) \, du = \frac{1}{\pi_{[b,a]}(h_0)} \int_0^\infty u \, J(a,u|h_0) \, du \tag{C.9}$$

By definition, the conditional probability density $T(a,x|b,0)$ of Eq. (27) is given by the limit of Eq. (C.8) for $h_0 \to b$, that is

$$T(a,x|b,0) = p_{[b,a]}(x|b) \tag{C.10}$$

and the function $X_T(a|b)$ defined by Eq. (27) is accordingly given by

$$X_T(a|b) = x_{[b,a]}(b) \tag{C.11}$$

By multiplying Eq.(C.3) by $x$, by integrating it from $x = 0$ to $x = \infty$, by taking into account that $J(a,0|h_0) = J(a,\infty|h_0) = 0$, and by making use of Eq. (C.4), one finds that the function

$$f_{[b,a]}(h_0) = \pi_{[b,a]}(h_0) x_{[b,a]}(h_0) \tag{C.12}$$

obeys the differential equation

$$\frac{1}{2} B^2(h_0) \frac{d^2 f_{[b,a]}}{dh_0^2} + A(h_0) \frac{df_{[b,a]}}{dh_0} + \pi_{[b,a]}(h_0) = 0 \tag{C.13}$$

with the boundary conditions

$$f_{[b,a]}(a) = f_{[b,a]}(b) = 0 \tag{C.14}$$

deriving from the fact that $x_{[b,a]}(a) = 0$ by definition, whereas $\pi_{[b,a]}(b) = 0$ because of Eq.(C.6). Equation (C.13) is a linear, non-homogeneous first-order differential equation for $df_{[b,a]}/dh_0$. The solution reads

$$\frac{df_{[b,a]}}{dh_0} = \psi(h_0) \left[ C - 2 \int_b^{h_0} \frac{\pi_{[b,a]}(u)}{B^2(u)\psi(u)} \, du \right] \tag{C.15}$$

where $\pi_{[b,a]}(u)$ and $\psi(u)$ are given by Eq.(C.7) and Eq. (32), respectively. The constant of integration $C$ can be expressed as

$$C = \frac{1}{\psi(b)} \left[ \frac{df_{[b,a]}}{dh_0} \right]_{h_0=b} \tag{C.16}$$

By deriving Eq.(C.12) with respect to $h_0$, by taking into account that $\pi_{[b,a]}(b) = 0$ and $[d\pi_{[b,a]}/dh_0]_b = \psi(b)/K_{[b,a]}$, and by making use of Eq.(C.11), we obtain

$$\left[ \frac{df_{[b,a]}}{dh_0} \right]_{h_0=b} = \frac{\psi(b)}{K_{[b,a]}} x_{[b,a]}(b) = \frac{\psi(b)}{K_{[b,a]}} X_T(a\,|\,b) \tag{C.17}$$

Equations (C.16) and (C.17) permit one to write Eq.(C.15) in the form

$$\frac{df_{[b,a]}}{dh_0} = \psi(h_0) \left[ \frac{X_T(a\,|\,b)}{K_{[b,a]}} - 2\int_b^{h_0} \frac{\pi_{[b,a]}(u)}{B^2(u)\psi(u)}\,du \right] \tag{C.18}$$

By integrating Eq.(C.18) from $b$ to $h_0$ and by inverting the order of integration in the double integral, one obtains

$$f_{[b,a]}(h_0) = X_T(a\,|\,b)\pi_{[b,a]}(h_0) - \frac{2}{K_{[b,a]}} \int_b^{h_0} \left[ \int_b^u \psi(u')du' \right] \left[ \int_u^{h_0} \psi(u')du' \right] \frac{du}{B^2(u)\psi(u)} \tag{C.19}$$

where use has been made of Eq.(C.7). Taking into account that $f_{[b,a]}(a) = 0$ and $\pi_{[b,a]}(a) = 1$, one concludes that $X_T(a|b)$ must be equal to Eq. (33).

[22] A possible way to give mathematical meaning to these considerations is to represent the action of $T_{[g]}(h)$ by a graph, where each point represents a certain strongly stable sub-interval of $[x_L, x_U]$, and the connecting arrows represent the various possible Barkhausen jumps from one sub-interval to another. Examination of this graph permits one to identify the weakly stable and the weakly unstable portions of the generating

function, and, on this basis, classify the various admissible jump structures relevant to hysteresis The analysis of these aspects is in progress and will be the subject of a future paper.

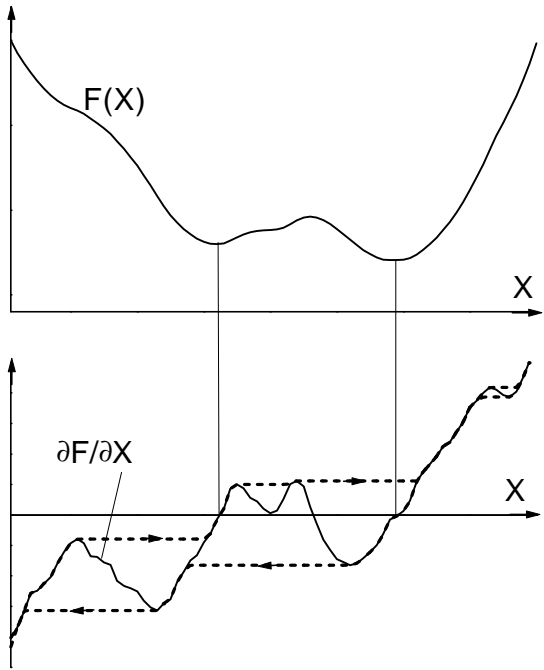

Fig. 1 – Free energy $F(X)$ with multiple minima and corresponding gradient $\partial F/\partial X$. The dashed line represents the hysteretic behavior of $H(X)$ obtained from the stability condition $H = \partial F/\partial X$.

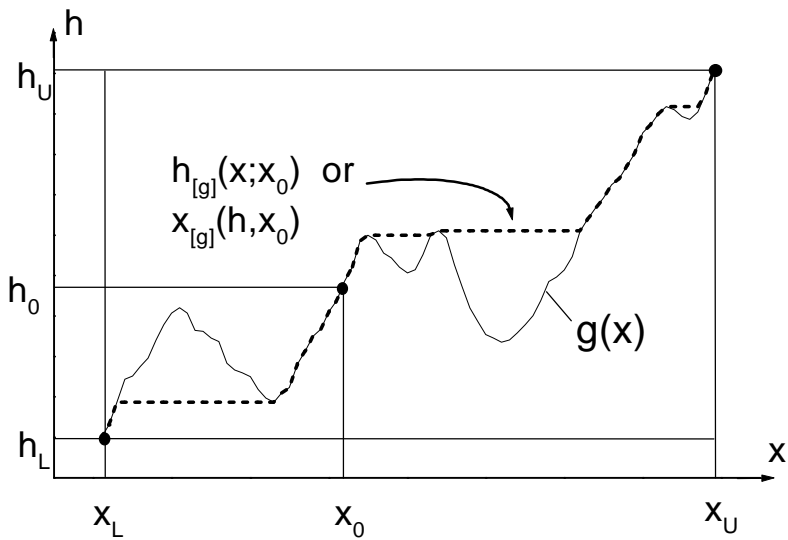

Fig. 2 – Generating function $g(x)$ with illustration of the envelope construction (broken line) associated with the function $h_{[g]}(x;x_0)$ and its inverse $x_{[g]}(h;x_0)$.

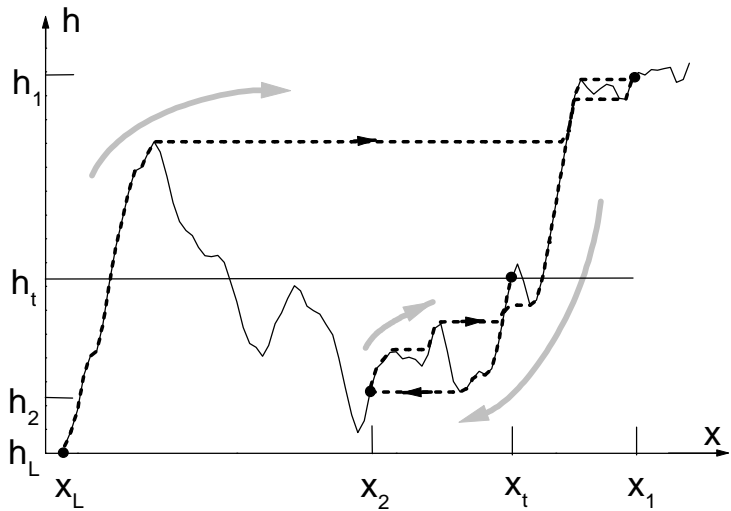

Fig. 3 – Action of evolution operators $T_{[g]}(h)$ associated with a given sequence of input reversals. Initial state is lower saturation $s_L = (h_L, x_L)$. Reversals take place at $s_1 = T_{[g]}(h_1)\, s_L = (h_1, x_{[g]}(h_1; x_L))$, $s_2 = T_{[g]}(h_2)\, s_1 = (h_2, x_{[g]}(h_2; x_1))$, and so on.

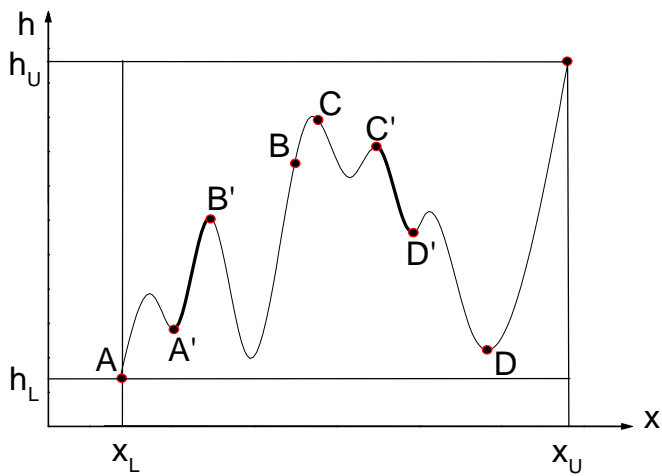

Fig. 4 – Illustration of intervals with different kinds of stability. AB: weakly stable; A′B′: strongly stable; CD: weakly unstable; C′D′: strongly unstable.

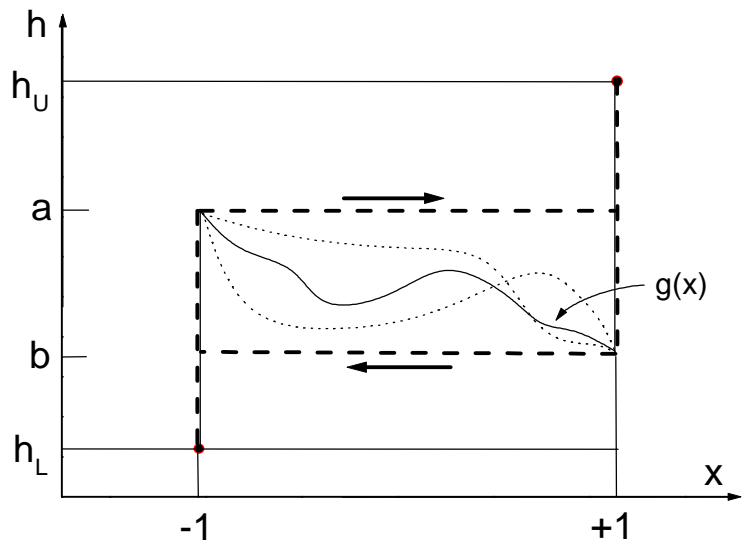

Fig. 5 – Typical generating function associated with the Preisach model (solid line), and corresponding envelope construction (dashed line, see also Fig. 3). The dotted lines give examples of different weakly unstable behaviors giving rise to the same hysteresis properties.

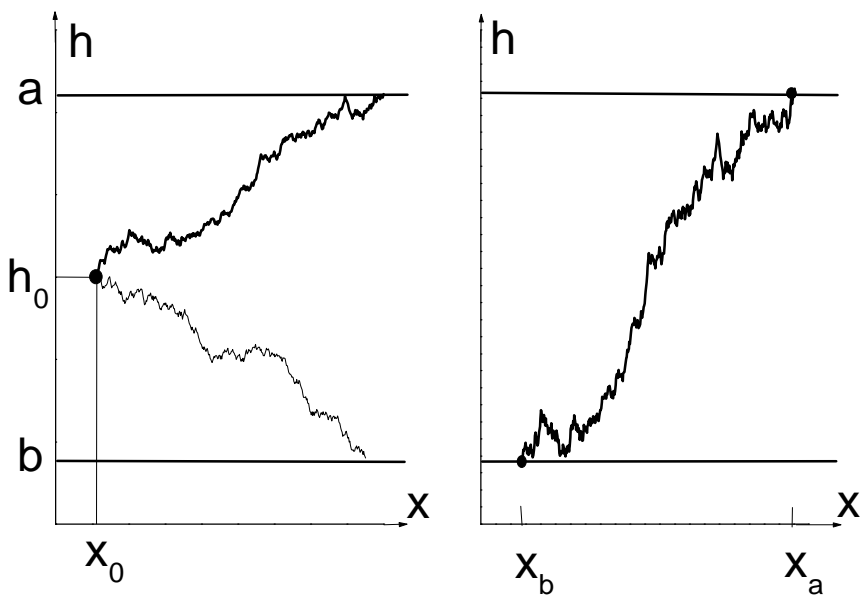

Fig. 6 – Left: example of stochastic process sample functions involved in the study of level-crossing through the boundary $h = a$ (thick line) or $h = b$ (thin line), starting from the initial condition $x = x_0$ at $h = h_0$. Right: same as before, in the particular case where $h_0 \to b$ and crossing through $h = a$ only is considered.

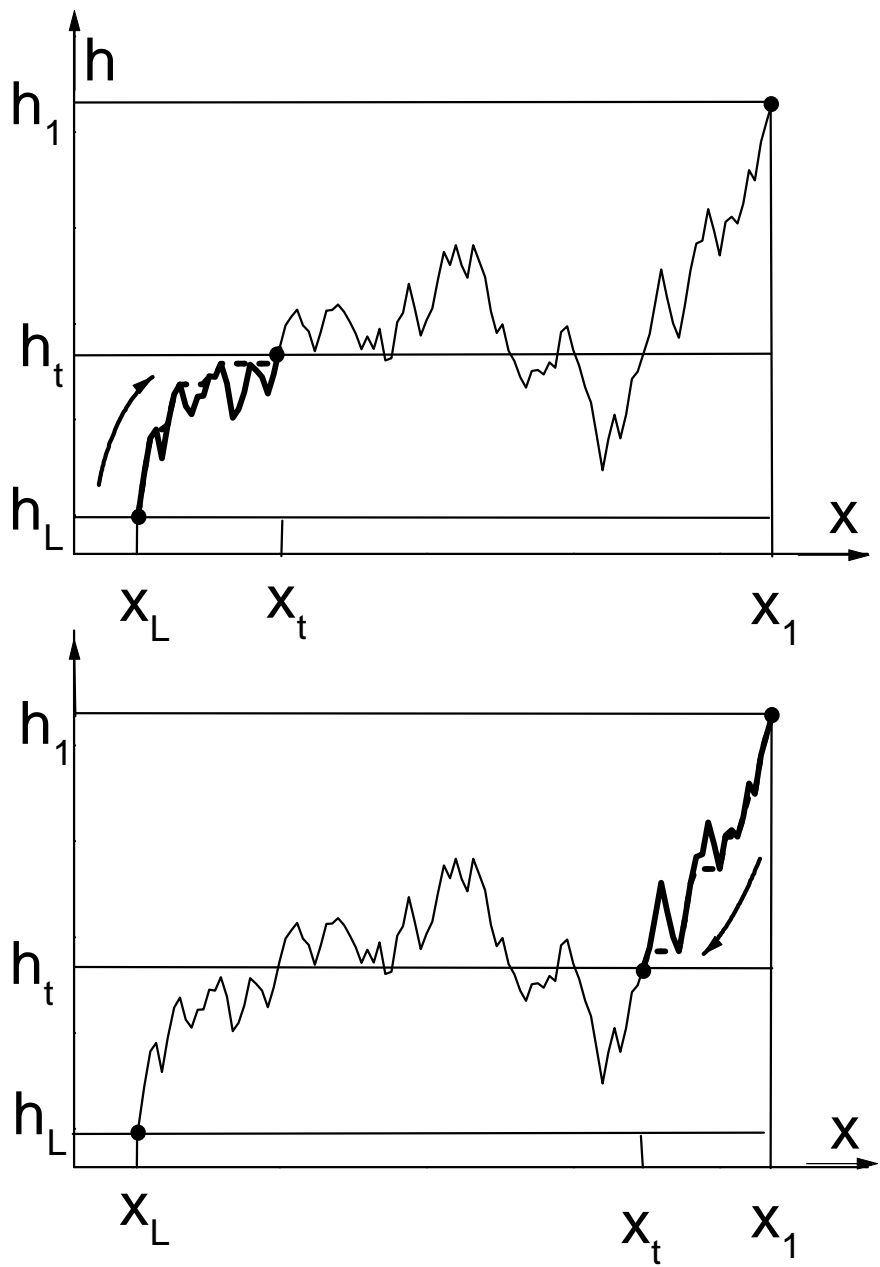

Fig. 7 – Level-crossing problems to be solved to calculate hysteresis in Markovian process. Top (field increasing from $h_L$ toward $h_1$): level-crossing must be considered in the interval $[h_L,h_t]$, with known distribution $p_L(x_L)$ at the *lower* h boundary $h_L$. Bottom (field decreases from $h_1$ toward $h_2$, not shown): level-crossing must be considered in the interval $[h_t,h_1]$, with known distribution $p_1(x_1;h_1)$ at the *upper* h boundary $h_1$.

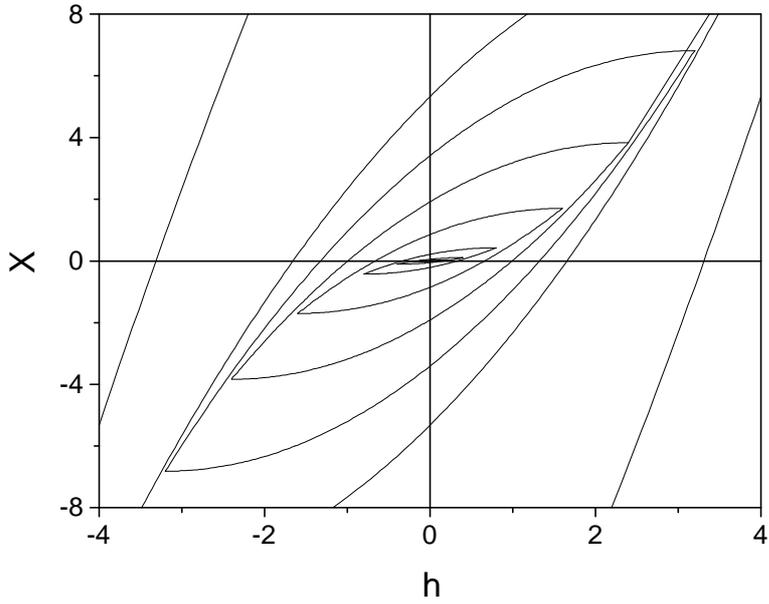

Fig. 8 – Typical hysteresis curves for Wiener process, calculated from Eq. (39).

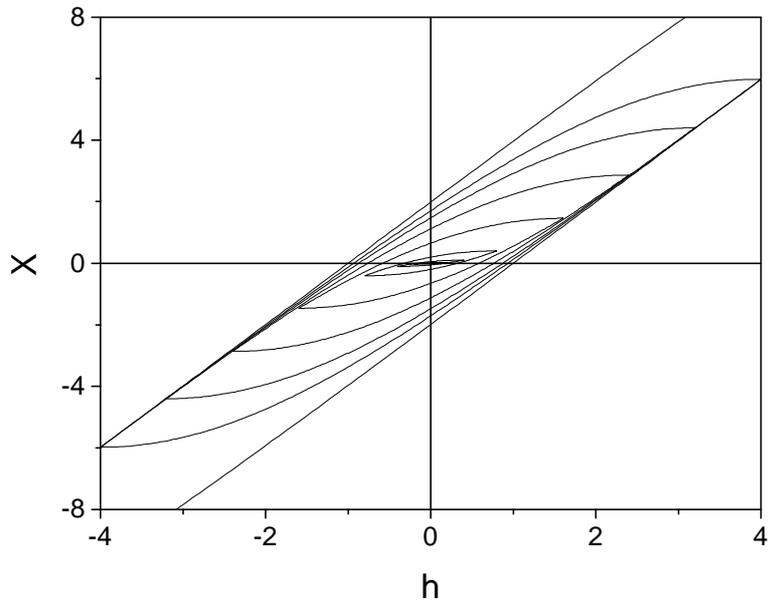

Fig. 9 – Typical hysteresis curves for Wiener process with drift, calculated from Eq. (41).